\documentclass[aps,apl,twocolumn,superscriptaddress,amsmath,amssymb,showpacs]{revtex4-1} 

\usepackage{color}
\usepackage{graphics}
\usepackage{bm}
\usepackage{epstopdf}
\usepackage{epsfig}

\begin{document}

\title{Probing the three-dimensional strain inhomogeneity and equilibrium elastic properties of single crystal Ni nanowires }
 
\author{E.~Fohtung}
\email[]{efohtung@physics.ucsd.edu}
\affiliation{ Department of Physics, University of California-San Diego, La Jolla, California 92093-0319, USA}
\author{J.~W.~Kim}
\affiliation{ Department of Physics, University of California-San Diego, La Jolla, California 92093-0319, USA}
\author{Keith T. Chan}
\affiliation{ Center for Magnetic Recording Research, University of California-San Diego, La Jolla, California 92093-0401, USA}
\author{Ross Harder}
\affiliation{ Advanced Photon Source, Argonne, Illinois 60439, USA}
\author{Eric ~E.~Fullerton}
\affiliation{ Center for Magnetic Recording Research, University of California-San Diego, La Jolla, California 92093-0401, USA}
\author{O.~G.~Shpyrko}
\affiliation{ Department of Physics, University of California-San Diego, La Jolla, California 92093-0319, USA}

\date{\today}
\begin{abstract}
\noindent We employ three dimensional x -ray coherent diffraction imaging to map the lattice strain distribution, and to probe the elastic properties of  a single crystalline Ni (001) nanowire grown vertically on an amorphous ${\text{Si}}{{\text{0}}_{\text{2}}}\parallel {\text{Si}}$ substrate. The reconstructed density maps show that with increasing  wire width, the equilibrium  compressive stress in the core region decreases sharply  while the surface tensile strain increases, and gradually trends to a nonzero constant. We use the retrieved projection of lattice distortion to predict the Young's Modulus of the wire based on the elasticity theory.
\end{abstract}

\maketitle

Due to their high surface-to-volume ratio, transition metal nanostructures such as nanowires (NWs) could potentially be used in a broad range of applications in catalysis, sensors, batteries, fuel cells, and magnetic devices.~\cite{huber2003raney,yang2011promotion,chan2007high,shibli2006nano}  Most of these devices require NWs with well-defined size, shape,  and spatial ordering. In addition, economical routes to mass production of NWs are desirable for practical applications. However, structural morphology of the NWs is strongly affected by the complex interplay between several growth parameters, such as temperature of the substrate, vertical and lateral material transfer, growth rate, in-plane mobility of ad-atoms, etc. In order to understand resulting morphology of NWs it is therefore vital to develop characterization tools that can nondestructively probe the three dimensional (3D) structural and mechanical properties of these NWs with nanoscale resolution.

\begin{figure*}[htb!] 
\resizebox{17.0cm}{6.50cm}{\includegraphics{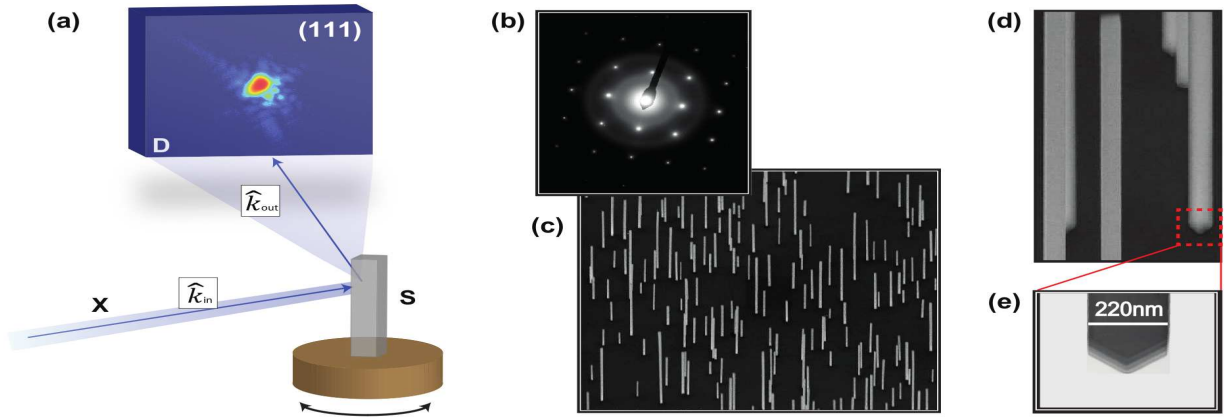}} 
\caption{\label{fig_setup}(Color online) Schematic of the experimental set-up and sample. (a) Monochromatic X-ray beam (X) of wave vector ${\widehat k_{in}}$ impinges on to Kirkpatrick-Baez  mirrors (not shown on sketch), which creates a localized illumination on the sample (S). By rotating the sample through the Bragg condition in increments of about 0.005 degrees, CXD patterns in the vicinity of the (111) reciprocal lattice point are recorded with a two-dimensional pixelated detector D. A typical diffraction pattern (out of the hundreds of patterns collected) is shown here. (b)  Selected area electron diffraction pattern from a vertical Ni NW indicates that it is a single crystal with a 001 oriented growth axis. (c) and (d) Scanning electron micrographs of the sample reveal that the NWs are well oriented with smooth facets. (e) Basal part of the NW exhibits  pronounced tapering features and orthogonal faceting. }
\end{figure*}

In this letter we study the structure, shape, 3D lattice distortion, and the electron density distribution in single crystal Ni NWs with the aid of synchrotron-based coherent x-ray diffraction (CXD) imaging  \cite{pfeifer06, robinson09}.   The NWs were grown using thermal chemical vapor deposition (CVD) on a Si substrate that has been oxidized resulting in a 500-${\text{nm}}$ SiO$_2$ amorphous coating layer. CVD at ${\text{650}}{{\text{ }}^{\text{o}}}{\text{C}}$  yields densely populated coverage of vertically-oriented single-crystal NWs depicted in Fig.~\ref{fig_setup}(b) with well-defined orthogonal  and smooth facets.~\cite{chan2010oriented,chan2012Philosophical}  
The NWs are grown in the [001] crystallographic orientation with  widths ranging from 50 ${\text{nm}}$ to about 300 ${\text{nm}}$ and lengths of up to 5 $\mu m$ and average coverage density  of between 0.1 and 0.3  ${\text{NW/}}\mu {{\text{m}}^2}$.  

CXD in the Bragg geometry has been shown to be a powerful characterization technique for imaging of local nanoscale lattice distortions, commonly described by deformation fields, ${\bm u}({\bm r})$ within a small crystal.  \cite{pfeifer06, robinson09} CXD allows us to image not only the overall shape of the nanostructure in 3D, but also the projection of the crystal lattice displacement field on to the Q vector of the measured Bragg spot. The lattice distortion represents the imaginary component of object density and manifests itself as local asymmetry of the measured Bragg diffraction pattern. Upon an inversion of coherent x-ray diffraction pattern, the resulting real-space image of the object will be complex valued $\widetilde \rho \left( {\bf{r}} \right) = {\rho _{{{\bf{G}}_{hkl}}}}\left( {\bf{r}} \right)\exp \left[ { - i{{\bf{G}}_{hkl}} \cdot {\bf{u}}\left( {\bf{r}} \right)} \right]$, where the amplitude ${\rho _{{{\bf{G}}_{hkl}}}}$ represents the density of the object and the phase, ${{\bf{G}}_{hkl}} \cdot {\bf{u}}\left( {\bf{r}} \right)$  is the  projection of the atomic displacement vector field ${\bf{u}}\left( {\bf{r}} \right)$  relative to atomic positions in an ideal perfectly periodic lattice onto the reciprocal lattice vector ${{\bf{G}}_{hkl}}$ for the measured Bragg reflection ($hkl$). Since only the intensities, and not the phases, are measured experimentally, direct inversion of  the CXD pattern to obtain  a real-space distribution of the object density is impossible due to phase problem.  However, the phase problem~ \cite{Fienup93} can be solved via special phasing procedures,  provided that the scattering object is finite and that diffraction pattern is sufficiently oversampled. ~\cite {miao2000oversampling,miao1999extending,pfeifer06,Chapman:06,minkevich2011strain,robinson09}

Figure~\ref{fig_setup}(a) shows the experimental setup used for CXD at Beamline 34-ID-C at the Advanced Photon Source (Argonne National Laboratory). A monochromator was used to select $E = 8.919$~keV x-rays with 1~eV bandwidth, resulting in a longitudinal coherence length of about $0.7\mu m$.  Slits were placed into the beam upstream of the Kirkpatrick-Baez focusing mirrors to select a $20\mu m \times 50\mu m$ wide transversely coherent segment of the beam that was focused down to about $1.5\mu m$ to illuminate a single NW. Coherent diffraction patterns were recorded for the rocking curves of the (004) and (111) Bragg reflections by rotating the sample through the Bragg condition in increments of about 0.005 degrees.  A CCD detector with $20\mu m$ pixel size was used to collect the 2D diffraction slice for each rotation angle. Full 3D diffraction patterns were then constructed by tomographically stacking these 2D frames together. 

\begin{figure}[t] %
\centering
\includegraphics[width=7.50cm]{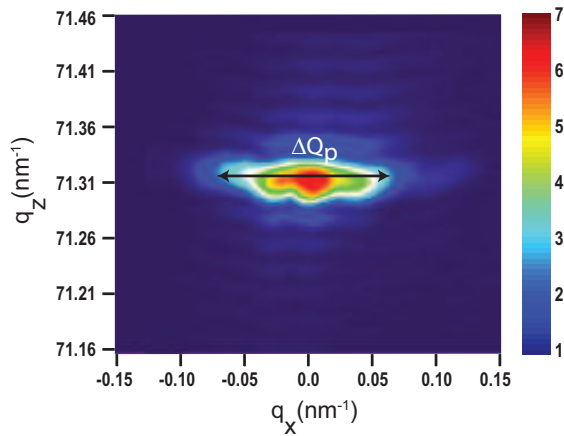} 
\caption{(Color online) Measured reciprocal space map (log 10 of intensities) from Ni NW  near the 004 Bragg reflection.}
\label{RSM}
\end{figure}

The lack of inversion symmetry for CXD patterns such as the one shown in Fig.\ref{fig_setup}(a) and Fig.\ref{RSM}  are a clear indication that the imaged nanostructure is strained, resulting in non-zero imaginary component of effective density.\cite{harder2010imaging} We inverted our CXD patterns using a combination of Fienup's Hybrid Input-Output~ \cite{Fienup93} and Error Reduction algorithms. Electron diffraction patterns (see Fig.~\ref{fig_setup}(b)) and SEM (Fig.~\ref{fig_setup}(c-d)) indicate that the NWs are defect-free, vertically oriented (grown along 001) single crystals with smooth facets.  A priori knowledge of the NWs shapes from SEM and the thickness oscillations along the vertical direction clearly seen from the diffraction patterns in Fig.~\ref{fig_setup}(a)  and  Fig.~\ref{RSM} were utilized to determine a good starting guess of the size of the region (support) at which the reconstructed object is allowed to exist. 

From a priori NWs growth information,~\cite{chan2010oriented, chan2012Philosophical}  we observed that the NWs had a small amount of impurities embedded at the surface. Taking this into account, we introduce an inner \textit{core} region  of the NW to be of uniform density and having relatively few impurities while the outer \textit{shell} region has a higher density of impurities. The x-ray diffraction patterns from this core-shell  model are composed of an interference sum of individual wave field contributions scattered from the two regions: core and shell of the NW. We first define an un-perturbed state of the crystal by assuming that for each region there is idealized homogeneous composition and strain, spatially limited by the shapes of the individual crystals. We then introduce the perturbation by taking into account the effects of non-uniform strain relaxation and possible compositional fluctuations within the previously homogeneous region of the crystals. The contribution of each wave field is the result of the convolution of two Fourier transforms: (1) the individual reciprocal shape function which leads to intensity decay of at least  ${q^{ - 2}}$ with respect to the unperturbed Bragg peak of that material and (2) the Fourier transform of $\widetilde \rho \left( {\bf{r}} \right)$ containing the spatial fluctuations in amplitude (compositional fluctuation) and phase (displacement field). 

The spread of the latter Fourier transform in reciprocal space  (see Fig.~\ref{RSM}) is dependent on the maximum values of the spatial gradients of the displacement field in the idealized layers and is related to the strain-induced broadening of the diffracted signal $\Delta {Q_p}$  (along a given $p$ direction)  near the Bragg spot. A more detailed description of this approach can be found in our earlier work.  \cite{minkevich2011strain, minkevich2009selective} We transform this Òband-limitedÓ frequency domain constraint into a direct space constraint of limited lattice displacement gradient.  By utilizing the uniformity of the electron density (and its threshold) constraints \cite{minkevich2009selective}  along with  the standard support constraints, we reconstruct the 3D distribution of NW electron density (shape) and corresponding phases (related to the projection of displacement field) from the (004) and (111) Bragg peaks, shown in Figs.~\ref{fig_setup}a). 
We estimate the magnitude of the maximum values of displacement derivative along the vertical direction (from 004  Bragg peak) and along the (111) direction (from 001 Bragg peak). These values were used as a limit for the maximum allowable phase difference in real space for neighboring points in the axial and the (111) directions. 
 
\begin{figure}[t]%
\centering
\includegraphics[width=7.60cm]{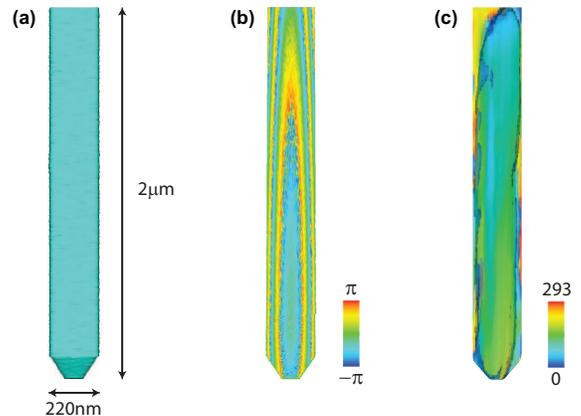}
\caption{(Color online) 3D reconstructions of (a) support, (b ) phases  and (c) amplitudes of the NW complex valued density function from the (004) Bragg peak.}
\label{edensity}
\end{figure}

\begin{figure}
  \centerline{\includegraphics [clip,width=3.60in, angle=0]{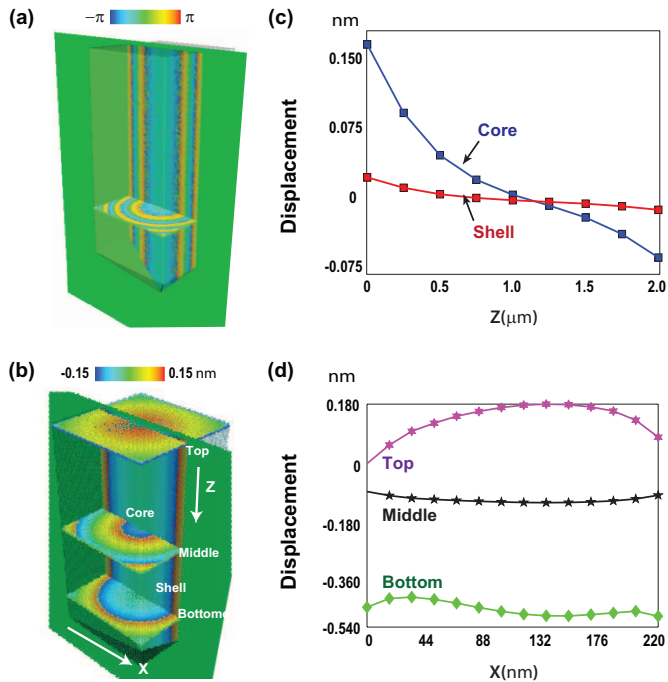}}
  \caption{(Color online) 3D reconstructions of (a) the phases, and (b) the un-wrapped phases from (004)  and (111) Bragg peaks for a single crystal Ni NW converted to projection of  the vertical  and axial components of the displacement field. (c) The variation of vertical component of displacement field along the axial (001) direction within the core (blue) and the shell (red) of the NW. (d) The variation of the vertical component of displacement field along the radial (110) direction of NW width at the top (purple), middle (black) and bottom (green) parts of the NW, at select regions indicated in b).}
  \label{fig_phase}
\end{figure}

 Figures~\ref{edensity}(a) and \ref{edensity}(b) show the  refined support (outer shape) and  the reconstructed phases of the NW while the  reconstructed amplitudes (internal distribution of electron density) are displayed in  Fig.~\ref{edensity}(c). The solution to our phasing problem puts the NW width at about $220~nm$ and length of $2~\mu m$, consistent with SEM measurements. A 3D reconstruction of the distribution of phases from the (004) and (111) Bragg peaks are shown in Fig.~\ref{fig_phase}(a). Unwrapping these phase distributions gives us the magnitude of the axial (001) and radial (110) components of the displacement field Fig.~\ref{fig_phase}(b). The color-bar shows that the maximum average displacement of 0.15 nm (0.4 of the Ni lattice parameter) is attained at the top of the NW.  The fact that the bottom of this NW is contracted may be attributed to the formation of free surfaces from the NW tapering. The height of the tapered section will depend on the specific sample, but the angle between the main body and the tapering portion could be formed by the $(110)$ and $(100)$ planes.~\cite{chan2010oriented, chan2012Philosophical} To test this assumption, we utilized the Bragg peak broadening in the vicinity of (111) and (004) to obtain a measure of the maximum displacement gradient along those crystallographic directions. \cite{minkevich2009selective,minkevich2011strain} 
 
 In order to confirm the reproducibility and uniqueness of the obtained solutions, we performed a series of phase retrieval procedures with different input random set of phases for the measured scattered radiation. We defined a cost function in the form of an error metric as $E^2_k = {\sum^{N}_{i=1} ( |F^{sim}_i|- \sqrt{I^{Exp}_i} )^2}/{\sum^{N}_{i=1} I^{Exp}_i}$, where $|F^{sim}_i|$ is the magnitude of the simulated amplitude and $I^{Exp}_i$ is the experimental intensity of point $i$ in the reciprocal space map. 

The phasing process, for a given initialization of the algorithm, was determined to be complete when the error metric is reduced to $10^{-8}$. The maximum number of iterations allowed was typically about 4,000. The differences between many equivalent solutions obtained from random initial sets phases provide a measure of the resolution of the resulting NW reconstruction in the form of a Phase Retrieval Transfer Function.\cite{ Chapman:06} The real-space resolution attained in this work is about $20\times 20$~nm$^2$.  
 
From Figs.~\ref{fig_phase}(c) and \ref{fig_phase}(d) we notice the presence of  a tensile strain at the near-surface regions and at the tip of the NW, contrasted with significant compressive  strain in the core. We observe that the tensile strain decays non-uniformly along the $z$-axis while changing smoothly in the core region to a compressive strain at the bottom of the NW. At the top-shell region of the NW (see Fig.~\ref{fig_phase}(d)), the strain is purely tensile but nonzero.  We also notice (see Fig.~\ref{fig_phase}(c)) that the core region of the NW has a resultant uniaxial compressive lateral displacement field ranging from zero  to  about  -0.15~nm while the maximum tensile displacement magnitude is 0.15~nm. The core region of the NW shows a clear homogenous density distribution while at the shell it varies slightly, as shown in Fig.~\ref{edensity}(c). This can be attributed to the presence of surface impurities (likely Si) during NW growth. These impurities are associated with a slight change in the average electron density.  This causes a change of the optical path length of the x rays and thereby a phase variation. However, this phase information alone is insufficient to draw definitive conclusions on the quantitative impurity contents.

 From the retrieved displacements, utilizing the elasticity theory \cite{landau1982theory} and experimentally verified values for the stiffness constants,  we obtain quantitative change in the stress from $-3.6~{\text{ GPa}}$ at the center of the NW to about $13.4~{\text{ GPa}}$ in the outermost part. However, the stress remains approximately constant within a  core region of the NW where the density is highly uniform. This nonzero stress is due to residual strain in the NW.. This implies that the surface endures the tensile strain while the interior is compressive, even if this NW is in equilibrium, thus providing a possible explanation for observed tapering. 
 
 Most metals have the possibility of exhibiting nonlinear and yet elastic behavior i.e., the stress $\sigma$ is essentially nonlinear to strain within the elastic limit.~\cite{landau1982theory} Under such conditions, the stress-strain relation can be approximated by a Taylor series expansion $\sigma  = {\sigma _0} + {a_1}\varepsilon  + {a_2}{\varepsilon ^2} + 0\left( {{\varepsilon ^3}} \right) \approx {a_1}\varepsilon  + {a_2}{\varepsilon ^2}$. The term ${\sigma _0}$ is negligible if the crystal is  unstrained. For extremely small strain, i.e., if the length-scale over which the deformation varies is much larger than the discrete length of the matter ~\cite{landau1982theory} (in this case much larger than the lattice parameter of Ni), the stress functions obey a quadratic strain behavior.  By adopting this quadratic type variation, we obtain the quantitative value of the Young's modulus of Ni NW given by  ${a_1}$ to be $100~{\text{ GPa}}$ while its dependence due to strain given by ${a_2}$ is $480~{\text{ GPa}}$.
 
In conclusion, we have used CXD to investigate the shape and internal structure of CVD-grown single crystal Ni NW grown on an amorphous ${\text{Si}}{{\text{O}}_{\text{2}}}$ substrate. We demonstrate that the basal NW taper is accompanied by substantial residual strain within the NW, as evidenced by a transition from compressive strain in the core of the NW to tensile strain in the near-surface shell. The quantitative values of the elastic properties such as the Young's Modulus and its dependence on strain have been calculated from the imaged strain field using the elasticity theory. Characterization of the Ni NWs confirms their exceptional crystalline quality and indicates their suitability for numerous wide-ranging applications, while strain engineering opens a gateway to novel NW-based device functionalities under external stimuli such as pressure, electric or magnetic fields.

Use of the Advanced Photon Source is supported by the US Department of Energy, Office of Science, Office of Basic Energy Sciences, under Contract DE-AC02-06CH11357. EF, JWK and OGS were supported by US Department of Energy, Office of Science, Office of Basic Energy Sciences, under Contract DE-SC0001805. KTC and EEF were supported by NSF Award DMR-0906957.

\end{document}